\documentclass[12pt]{amsart}

\usepackage{amsmath, amsthm, amscd, amsfonts, amssymb}
\usepackage{txfonts} 
\usepackage{mathrsfs} 
\usepackage{textcomp}
\usepackage{graphicx}
\usepackage{color}
\usepackage[colorlinks]{hyperref}
\usepackage[all]{xy}

\title[Causally simple spacetimes and Naked singularities]{Causally simple spacetimes and Naked singularities}
\author{Mehdi Vatandoost $^{1}$, Rahimeh Pourkhandani$^{2}$ and Neda Ebrahimi $^{3}$}
\email{m.vatandoost@hsu.ac.ir}

\email{r.pourkhandani@hsu.ac.ir}

\email{n$\underline{  \:\:}$ebrahimi@uk.ac.ir}

\address{$ ^{1, 2}$Department of Mathematics and Computer Sciences, Hakim Sabzevari University,
Sabzevar, Iran.}
\address{$ ^{3}$ Department of Pure Mathematics, Faculty of Mathematics and Computer, Mahani Mathematical 
Research Center, Shahid Bahonar University of Kerman, Kerman, Iran.}
\newtheorem{theorem}{Theorem}[section]
\newtheorem{lemma}[theorem]{Lemma}

\newtheorem{proposition}[theorem]{Proposition}
\newtheorem{corollary}[theorem]{Corollary}
\theoremstyle{definition}
\newtheorem{definition}[theorem]{Definition}

\theoremstyle{example}

\theoremstyle{remark}
\newtheorem{remark}[theorem]{Remark}

\newtheorem{My Guess}[theorem]{My Guess}
\newtheorem{conjecture}{Conjecture}
\begin{document}

\begin{abstract}
In this paper, We prove a conjecture which states that if $M$ is a nakedly singular future boundary or 
nakedly singular past boundary spacetime, then the space of null geodesics, $\mathcal{N}$, is non-Hausdorff.
Also, we show that every two-dimensional strongly causal spacetime $M$ is causally simple if and only if 
it is null pseudoconvex. As a result, it implies the converse of the conjecture for two-dimension but there are examples that refute it 
for more dimensions.
\end{abstract}

\maketitle
\section{Introduction and preliminaries}
%
The cosmic censorship hypothesis of Roger Penrose is one of the most outstanding conjectures in the gravitational theory to find the final fate of
the gravitational collapse of a massive star.
According to a given observer, this conjecture asserts there can be no naked singularity in spacetime or in other words, there exist no families 
of future directed causal curves, which in the past terminate at the singularity \cite{Joshi}.\\
The global hyperbolicity condition on a spacetime implies the existence of maximal length geodesics 
joining causally related points and so, it is generally believed that a mathematical statement of the cosmic 
censorship hypothesis is that spacetime should be globally hyperbolic. However, it seems that there are weaker 
conditions that may play the role of global hyperbolicity \cite{Beem(1987)}. One such condition is pseudoconvexity. 
The concepts of causal or null (or maximally null) pseudoconvexity are defined, by restricting the condition of 
pseudoconvexity to causal or null (or maximal null) geodesics, respectively. Various implications of causal and 
null pseudoconvexity on the geodesic structure of a Lorentzian manifold have been studied in several classical and recent papers 
by Beem, Parker, Krolak, and Low \cite{Beem(1987), Beem(1992), Beem(1996), Low(1989), Low(1990)}.
In Ref. \cite{Low(1990)}, Low states the equivalence of the null pseudoconvexity of $M$ and the Hausdorffness of the space of null 
geodesics, $\mathcal{N}$, 
for a strongly causal spacetime $M$. A sufficient condition to ensure that $\mathcal{N}$ is Hausdorff is the 
absence of naked singularities \cite[Proposition 2.2]{Low(1989)}. But, we can see in \cite[Example 2.2.12]{Bautista(2016)} 
that it is not a necessary condition.
Recently, Borjian and Bahrampour introduced two types of naked 
singularities called nakedly singular future boundary and nakedly singular past boundary \cite{Borjian}.\\
In this paper, by an example, we establish that a proposition proved by Beem and Krolak (see \cite [Proposition 1]{Beem(1992)}) 
is not valid and it is required to have a minor modification. Then, by the corrected version of this proposition,
we show that the existence of (at least) one of these naked singularities implies the failure of the Hausdorff property for 
the space of null geodesics of $M$ which is Conjecture 3.1 of \cite {Borjian} for causally continuous spacetime. 
Also, the converse of this conjecture is solved.\\
The causal ladder is a diagram that illustrates the strengths of the causality conditions (see \cite[Page 73, Fig. 3.3]{Beem(1996)} and 
\cite{Min(2008)}). 
As these various applications of pseudoconvexity show, it is useful to place this property within the causal ladder.
Recently, we proved that every strongly causal spacetime is causally simple if and only if it is maximal 
null pseudoconvex \cite{Vatan}. On the other hand, there is a conjecture that says strongly causal null 
pseudoconvex spacetimes are causally simple \cite{Bautista(2017)}. We also prove this conjecture 
for $n$-dimensional spacetimes ($n\geq 3$) and the converse of it for
two-dimensional spacetimes and then give a new hierarchy in the causal ladder (see Fig \ref{Fig1}).\\
Here, we briefly review some basic definitions and concepts from the topic of Lorentzian causality theory needed for the next sections.\\
%
%
In general relativity, a \textit{spacetime} is a pair $(M,g)$ where $M$ is a real, connected, $C^{\infty}$ Hausdorff manifold of dimension two 
or more, and $g$ is a globally defined $C^{\infty}$ Lorentzian metric on $M$ of signature $(+,-,...,-)$. When there is no ambiguity, we use $M$ 
to refer to the spacetime $(M,g)$.\\ 
We say that a vector $v\in T_{p}M$ is \textit
{timelike} if $g_{p}(v,v)>0$, \textit {causal} if 
$g_{p}(v,v)\geq0$, \textit {null}
if $g_{p}(v,v)=0$ and \textit {spacelike} if 
$g_{p}(v,v)<0$. 
A smooth curve is called \textit {future directed 
timelike curve} if its tangent vector is everywhere 
timelike future pointing vector and similarly for 
spacelike, causal, null future directed (or null past directed)
curve can be defined.
If $p, q \in M$, then $q$ is in the \textit
{chronological future of $p$}, written $q\in I^{+}(p)$ or $p \prec q$, 
if there is a timelike future pointing curve $\gamma: 
[0, 1]\rightarrow M$ with $\gamma(0) = p$, and 
$\gamma(1) = q$;
similarly, $q$ is in the \textit{causal future of $p,$}
written $q\in J^{+}(p)$ or $p \preceq q$, if there is a future pointing
causal curve from $p$ to $q$. For any point, $p$, the set $I^{+}(p)$ is open; 
 but $J^{+}(p)$ need not, in general, be closed. $J^{+}(p)$ 
 is always a subset of the closure of $I^{+}(p)$.
To be more careful, it is useful to recall that \textit {the causal ladder} is a set of conditions 
on spacetimes, where each of these implies its previous one \cite{Beem(1996)}:
\begin{enumerate}
\item A spacetime $M$ which has no point $p$ with 
a non-degenerate causal curve that starts and ends at $p$
is said to satisfy \textit {the causal condition}.
\item A spacetime $M$ is said to be distinguishing if for all points $p$ and $q$ in $M$, either 
$I^{+}(p) = I^{+}(q)$ or $I^{-}(p) = I^{-}(q)$ implies $p = q$.
\item If each point $p$ has arbitrarily small neighborhoods 
in which any causal curve intersects in a single component,
$M$ satisfies the condition of \textit {strong causality}.
\item A distinguishing spacetime $M$ is said to be \textit{causally continuous} at $p$ if the set-valued functions 
$I^{+}$ and $I^{-}$ are both inner continuous and outer continuous at $p$. 
The set-valued function $I^{\pm}$ is said to be \textit{inner continuous} at $p\in M$ if for each compact set 
$K \subseteq I^{\pm}(p)$, there exists a neighborhood $U(p)$ of $p$ such that $K \subseteq I^{\pm}(q)$ 
for each $q \in U(p)$. The set-valued function $I^{\pm}$ is \textit{outer continuous} at $p$ if for each compact set $K$ 
in the exterior of $\overline{I^{\pm}(p)}$ there exists some neighborhood $U(p)$ of $p$ such that for each $q \in U(p)$, $K$ 
is in the exterior of $\overline{I^{\pm}(q)}$. We recall that $I^{\pm}$ is always inner continuous (see \cite[Proposition 4.3]{Min(2019)}).
\item If $M$ is distinguishing and $J^{\pm}(p)$ is closed for all $p\in M$,
then $M$ is \textit{causally simple}.
\item A spacetime $M$ is said to be \textit{globally 
hyperbolic} if $M$ is strongly causal and $J^{+}(p)
\cap J^{-}(q)$ is compact for all $p$ and $q$ in $M$.
\end{enumerate}
\begin{definition}\label{def-maximal}
\cite{Beem(1992)} A future null geodesic ray $\gamma: [0,b) \rightarrow 
M$ is said to be \textit{maximal} if $\gamma (t) \not \in I^{+}
(\gamma(0))$ for all $t\in(0,b)$. In other words, if 
$t_{1}\neq t_{2}$ and $t_{1},t_{2}\in [0,b)$, 
then $\gamma( t_{1})$ and $\gamma( t_{2})$ are 
not chronologically related. Also, the cases $\gamma: [a,b] \rightarrow M$ and $\gamma:(a,b) \rightarrow M$ are 
defined as maximal null geodesic segment and maximal null geodesic in the same manner.
\end{definition}
A spacetime $(M, g)$ is said to be \textit{pseudoconvex} if and only if given any compact set $K$ in $M$, 
there exists always a larger compact set $K^{\ast}$ such that all ``geodesic segments'' joining points of 
$K$ lie entirely in $K^{\ast}$.
There are different types of pseudoconvexity which correspond to different classes of geodesics in spacetimes. 
For example, causal, null, and maximal null pseudoconvexity can be defined by restricting the condition ``geodesic segments''
to causal, null, and maximal null geodesic segments, respectively \cite{Beem(1987)}.
\begin{remark}\label{remark1}
Since every maximal null geodesic is a null geodesic and every null geodesic is a causal geodesic, it is clear from 
the definitions that causal pseudoconvexity implies null pseudoconvexity and 
null pseudoconvexity implies maximal null pseudoconvexity.
\end{remark}
The limit curve theorems are surely one of the most fundamental tools of Lorentzian geometry. Their importance is certainly 
superior to that of analogous results in Riemannian geometry because in Lorentzian manifolds the curves may have a causal 
character and, hence, it is particularly important to establish whether two points can be connected by a causal, a timelike, or a lightlike curve.
There are different forms of convergence for a sequence of 
nonspacelike curves $\{\gamma_{n}\}$ in Lorentzian geometry and general relativity. For example, the limit curve convergence, 
the $C^{0}$ convergence, and the uniform convergence. For arbitrary space-times, each of 
the limit curve convergence and the $C^{0}$ convergence is not stronger than the other. But in strongly causal space-times, 
these two types of convergence are almost equivalent for sequences of causal curves (see \cite[Proposition 3.34]{Beem(1996)}). 
Recently, Minguzzi introduced a discussion of the history of limit curve theorems 
results in Lorentzian geometry and proved a strong version of limit curve theorems by a generalized version of uniform convergence \cite{Min(2008-1)}.
\begin{definition}\label{def-uniform}
\cite[Definition 2.1]{Min(2008-1)} (In this definition $a_{n}, b_{n}, a,$ and $b$ may take an infinite value.)
Let $h$ be a Riemannian metric on $M$ and let $d_{0}$ be the associated Riemannian distance. The sequence of curves  
$\gamma_{n} : [a_{n},b_{n}] \rightarrow M$ converges $h$-uniformly to $\gamma : [a,b] \rightarrow M$ if 
$a_{n} \rightarrow a, b_{n} \rightarrow b,$ and for every $\epsilon > 0$ there
is $N > 0$, such that for $n > N$, and for every $t \in [a,b] \cap [a_{n},b_{n}], d_{0}(\gamma (t), \gamma_{n}(t)) <\epsilon $. 
\end{definition}
The sequence of curves $\gamma_{n} : [a_{n},b_{n}] \rightarrow M$ converges $h$-uniformly on compact subsets to 
$\gamma : [a,b] \rightarrow M$ if for every compact interval $[a^{\prime},b^{\prime}] \subseteq [a,b]$, there is a choice of sequences 
$a_{n}^{\prime},b_{n}^{\prime} \in [a_{n},b_{n}]$, $a_{n}^{\prime} < b_{n}^{\prime}$, such that $a_{n}^{\prime} \rightarrow a^{\prime}$, $b_{n}^{\prime} \rightarrow b^{\prime}$, and for any such choice $\gamma_{n} \vert _{[a_{n}^{\prime},b_{n}^{\prime}]}$ converges $h$-uniformly to $\gamma \vert _{[a^{\prime},b^{\prime}]}$.
Also, Minguzzi proved that the $h$-uniform convergence implies the $C^{0}$ convergence and on compact subsets, 
it is independent of the Riemannian metric $h$ chosen (see \cite[Theorem 2.4]{Min(2008-1)}). In this paper, by the 
limit curve or the limit geodesic segment, we mean that the $h$-uniform convergence on compact subsets is applied.\\
In Ref. \cite{Vatan}, we found an equivalent property to the pseudoconvexity, called the \textit{LGS property}:
%
%
\begin{definition}\label{def-LGS}
Assume $p_{n} \rightarrow p$ and $q_{n} \rightarrow q$ for distinct points $p$ and $q$ in a 
spacetime $M$. We say that the spacetime $M$ has the limit geodesic segment property (LGS), if 
each pair $p_{n}$ and $q_{n}$ can be joined by a ``geodesic segment'', then there is a limit geodesic 
segment from $p$ to $q$. Namely, for every sequence of geodesics $\gamma_{n}$ from $p_{n}$ to $q_{n}$ 
where $p_{n} \rightarrow p, q_{n} \rightarrow q,$ there are a subsequence $\gamma_{k}$ and a geodesic 
segment from $p$ to $q$ such that $\gamma_{k}$ converges $h$-uniformly to $\gamma$.
Similarly, causal, null, and maximal null LGS property can be defined by restricting the condition ``geodesic segment''
to causal, null, and maximal null geodesics, respectively.
\end{definition}
\begin{proposition}\label{causal LGS}
\cite[Proposition 4]{Vatan} Let $ (M, g)$ be a strongly causal spacetime. Then, it is (null or maximal null) causal pseudoconvex 
if and only if it has the (null or maximal null) causal LGS property.
\end{proposition}
%
%
%
%
%
%
\section{Main results}                                                                                                  %
%
The paper builds on the literature on pseudoconvexity and causality theory \cite{Beem(1987),Beem(1992), Beem(1996), Borjian, HOWKING(1973), 
Min(2019), Penrose, Vatan}. 
\subsection{Pseudoconvexity and causal simplicity conditions}                                       %
%
%

Null pseudoconvexity is a property weaker than causal pseudoconvexity. 
There are some examples of spacetimes which are null pseudoconvex but not causally pseudoconvex (see \cite[Example 1]{Vatan}).
%
\begin{figure}
\center
\vspace{0.1cm}
\hspace{2cm}
\includegraphics [width=10cm,height=10cm]{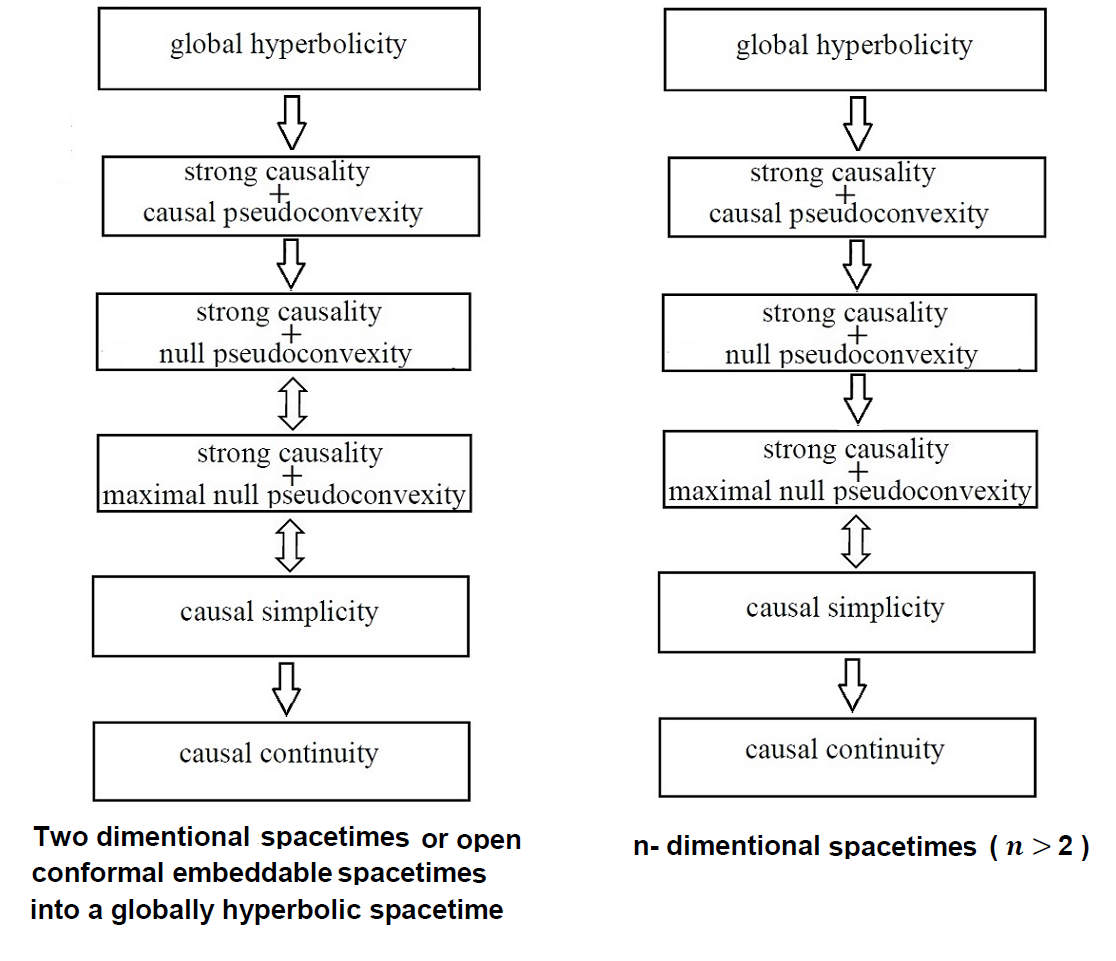}
\caption{A new hierarchy in the causal ladder by incorporating our new result on the three types of pseudoconvexity conditions.}
\label{Fig1}
\end{figure}
%
During the construction of a boundary for spacetimes, Bautista et al. studied in detail the space of light rays 
and conjectured that strongly causal null pseudoconvex spacetimes are causally simple \cite{Bautista(2017)}.
Beem and Krolak in \cite[Theorem 1]{Beem(1992)} proved that causal simplicity implies maximal null pseudoconvexity.
In Ref. \cite{Vatan}, we recently proved the converse of this fact in the case of strongly causal spacetime, that 
is a refined version of the conjecture.
\begin{theorem}\label{causally simple-null pseudo}
\cite[Theorem 2]{Vatan} Let $ (M, g)$ be a strongly causal spacetime. $(M,g)$ is causally simple if and only if it is maximally null pseudoconvex.
\end{theorem}
Also, Beem and Parker proved that global hyperbolicity implies causal pseudoconvexity (see \cite[Lemma 3.2]{Beem(1987)}). 
In Ref. \cite{Hedicke}, Hedicke and Suhr showed that the space of null geodesics of a n-dimensional causally simple spacetime $M$ for $n>2$ 
is Hausdorff if it admits an open conformal embedding into a globally hyperbolic spacetime. And by Theorem \ref{Low}, $M$ is null pseudoconvex.
%
Now, only one step remains to establish the ladder Fig \ref{Fig1}. Namely, in every two-dimensional strongly causal spacetime 
the maximal null pseudoconvexity implies the null pseudoconvexity. Before this, we need the following two lemmas that  
the proofs are straightforward.
\begin{lemma}\label{lemma seqen}
Let $M$ be a causally simple spacetime and $\{p_{n} \}$ and $\{q_{n}\}$ are sequences in $M$ converging to $p$ 
and $q$ respectively, and there are causal geodesic segments (even causal curves) $\gamma_{n}$ from $p_{n}$ to $q_{n}$ for all $n$. 
Then  $q \in J^{+}(p)$.
\end{lemma}
Every point $p$ in spacetime $M$ admits a local basis $\lbrace V_{k}, k \geq 1\rbrace,$ for the 
topology such that for every $k$, the open set $V_{k}$ is a relatively compact subset of $M$, a 
strictly convex normal neighborhood of $p$, and a globally hyperbolic spacetime itself
(see \cite[Theorems 1.35 and 2.7]{Min(2019)}).
Let $\gamma$ be a future directed null geodesic with past (future) endpoint $p$. In what follows, by a local extension of $\gamma$ at $p$,  
we mean the maximal extension of this geodesic into the past (future) in a strictly convex normal neighborhood of $p$.
\begin{lemma}\label{lemma order}
Let $(M, g)$ be a two-dimensional causally simple spacetime and let $p^{\prime}_{n} \rightarrow p$, $q^{\prime}_{n} \rightarrow q$, 
and $p^{\prime}_{n}$ can be joined to $q^{\prime}_{n}$ by a future directed null geodesic $\gamma^{\prime}_{n}$ for all $n$ and let
$\gamma_{n}$ be a local extension of $\gamma^{\prime}_{n}$ at $p^{\prime}_{n}$ and at $q^{\prime}_{n}$.
Then it is possible to choose a subsequence $\lbrace\gamma_{k}\rbrace$ of $\lbrace\gamma_{n}\rbrace$ and 
distinct points $p_{k}$ and $q_{k}$ on $\gamma_{k}$ for all $k$ such that $p_{k} \rightarrow p$, $q_{k} \rightarrow q$, and 
one of the following conditions is satisfied:
\begin{itemize}
\item[(1)] 
$p \prec  . . .  \prec p_{2}\prec p_{1} $ and $q \prec  . . . \prec  q_{2}\prec  q_{1}$. 
Obviously in this case, if $n>m$ then $p_{n}\not \in J^{+}(p_{m})$.
\item[(2)]
$p_{1}\prec p_{2}\prec . . . \prec p$ and $q_{1}\prec q_{2}\prec . . . \prec q$. 
Obviously in this case, if $n>m$ then $q_{n}\not \in J^{-}(q_{m})$.
\item[(3)]
The sequence $\lbrace\gamma_{k}\rbrace$ is constant.
\end{itemize}
Moreover, it is possible to choose the monotone sequences $\lbrace p_{k} \rbrace$ on a null geodesic passing through $p$ 
and $\lbrace q_{k} \rbrace$ on a null geodesic passing through $q$ (in the Cases (1) and (2) replace $\preceq$ with $\prec$).
\end{lemma}
%
\begin{remark}\label{remark2}
We note that any limit curve of a sequence of maximal null geodesics is a maximal null geodesic. 
For this, let $\gamma$ be a future directed causal curve from $p$ to $q$ as a limit curve of a sequence of 
future directed maximal null geodesics $\gamma_{n}$ from $p_{n}$ to $q_{n}$ such that $p_{k} \rightarrow p$ and $q_{k} \rightarrow q$. 
By the maximality of $\gamma_{n}$, the Lorentzian arc length of $\gamma_{n}$ is equal to the Lorentzian distance from $p_{n}$ to 
$q_{n}$ namely, $L(\gamma_{n})=d(p_{n},q_{n})$ (see \cite[Definitions 4.1 and 4.10]{Beem(1996)} 
that \cite[Definitions 4.10]{Beem(1996)} is equivalent to Definition \ref{def-maximal}). Now, By using \cite[Lemma 4.4]{Beem(1996)}, 
we have $L(\gamma) \leq d(p,q) \leq lim\, inf \, d(p_{n},q_{n})=0$. Hence $L(\gamma) = d(p,q)=0$ and $\gamma$ 
may be reparametrized to a maximal null geodesic segment from $p$ to $q$ by \cite[Theorem 4.13]{Beem(1996)}.
Because every null geodesic is locally maximal, then it immediately implies that any limit curve of a sequence of 
null geodesics is a null geodesic. For this, we can cover $\gamma$ with a finite number of strictly convex normal neighborhoods 
$U_{1}, ..., U_{m}$ such that $p\in U_{1}$, $q\in U_{m}$, and ${\gamma_{n}\vert}_{U_{i}}$ is a maximal null geodesic segment for all $i$. 
So, ${\gamma\vert }_{U_{i}}$ is a maximal null geodesic segment for all $i$.
\end{remark}
\begin{theorem}\label{main}
Let $M$ be a strongly causal two-dimensional spacetime. $M$ is maximal null pseudoconvex if and only if 
it is null pseudoconvex.
\end{theorem}
\begin{proof}
$(\Leftarrow)$ See Remark \ref{remark1}.
$(\Rightarrow)$ Suppose, on the contrary, $M$ is maximal null pseudoconvex but not null pseudoconvex.
Therefore, Theorem \ref{causally simple-null pseudo} implies $M$ is causally simple and Proposition \ref{causal LGS} 
implies the null LGS property is not satisfied. So, 
there are sequences $p_{n}$ and $q_{n}$ with $p_{n}\prec q_{n}$ joined by a future directed
null geodesic $\gamma_{n}$, for any natural value of $n$ without any limit curve from $p$ to $q$ 
(see Definitions \ref{def-uniform}, \ref{def-LGS}).
Let $h$ be a complete Riemannian metric on $M$. By \cite[Theorem 3.1, part (2), case (ii)]{Min(2008-1)}, 
there is a subsequence parametrized with respect to $h$-length denoted as $\gamma_{k}:[0,b_{k}]\rightarrow M$, 
$\gamma_{k}(0)=p_{k}\rightarrow p$, $\gamma_{k}(b_{k})=q_{k}\rightarrow q$, (an analogous reparametrized sequence 
$\gamma_{k}^{\prime}:[-b_{k},0]\rightarrow M$, $\gamma_{k}^{\prime}(0)=q_{k}\rightarrow q$, 
$\gamma_{k}^{\prime}(-b_{k})=p_{k}\rightarrow p$) and there are a future directed inextendible causal curve 
$\eta_{1}:[0,+\infty)\rightarrow M$, $\eta_{1}(0)=p$, and a past directed inextendible causal curve 
$\eta_{2}:(-\infty,0]\rightarrow M$, $\eta_{2}(0)=q$ such that $\gamma_{k}$ and $\gamma_{k}^{\prime}$ 
converges $h$-uniformly on compact subsets to $\eta_{1}$ and $\eta_{2}$, respectively. 
Now, Remark \ref{remark2} implies that $\eta_{1}$ and $\eta_{2}$ are null geodesics.
Also, we have $p\not\in \eta_{2}$ and $q\not\in \eta_{1}$; otherwise, $\eta_{1}$ and $\eta_{2}$ are one the reparametrization 
of the other from $p$ to $q$ and it leads us to get a contradiction (see \cite[Theorem 3.1, part (2), case (i)]{Min(2008-1)}).
Lemma \ref{lemma seqen} implies that $\overline{q} \in J^{+}(\overline{p})$ for 
all $\overline{p} \in \eta_{1}$ and $\overline{q} \in \eta_{2}$. 
We remark that if $\overline{q_0}=\eta_{2}(s_{0}) \in \partial J^{+}(p)$, then by \cite[Lemma 3]{Vatan}, 
$\eta_{2}(s)\in \partial J^{+}(p)$ for all real value $s\in (-\infty ,s_{0}]$). Now, one of the two cases occurs. 
Either there exists $\overline{p_{0}} \in \eta_{1}$ and $\overline{q_{0}} \in \eta_{2}$ such that 
$\overline{q_{0}} \in \partial J^{+}(\overline{p_{0}})$ and so $\eta_{2}((-\infty ,s_{0}])\subseteq \partial J^{+}(\overline{p_{0}})$, 
or for every $\overline{p} \in \eta_{1}$ and $\overline{q} \in \eta_{2}$, $\overline{q} \in I^{+}(\overline{p})$ 
and so $\eta_{2}\subset I^{+}(\overline{p})$ for all $\overline{p} \in \eta_{1}$.
\begin{itemize}
\item[\textbf{Case 1)}]
$\eta_{2}((-\infty ,s_{0}])\subseteq \partial J^{+}(\eta_{1}(t_{0}))$, for some $\overline{p_{0}} = \eta_{1}(t_{0})$ 
and $\overline{q_{0}} = \eta_{2}(s_{0})$. In this case, 
we consider 
$r= \eta_{2}(s_{1}) \in \partial J^{+}(\overline{p_{0}})$ for some $s_{1} \in (-\infty ,s_{0})$ and so by \cite[Corollary 4.14]{Beem(1996)}, 
there is a maximal null geodesic from $\overline{p_{0}}$ to $r$. 
By \cite[Proposition 2.19]{Penrose}, this yields one of the following two possibilities:\\
1-1) $\eta_{2}(s_{0})=\overline{q_{0}} \in I^{+}(\overline{p_{0}})$ which contradicts 
$ \eta_{2}(s_{0}) \in \partial J^{+}(\overline{p_{0}})$.\\ 
1-2) The union of segments $\overline{p_{0}}r $ and $r\overline{q_{0}} \subseteq \eta_{2}$ constitutes a single null geodesic from 
$\overline{p_{0}}$ to $\overline{q_{0}}$. Namely, $\overline{p_{0}}r \subseteq \eta_{2}$ and so $\eta_{1}$ and $\eta_{2}$ 
intersect each other at $\overline{p_{0}}$. 
By consideration of the two limit curves $\eta_{1}$ and $\eta_{2}$ of the sequence $\gamma_{k}$ in a strictly convex normal 
neighborhood $U(\overline{p_{0}})$ of $\overline{p_{0}}$, we conclude that $\eta_{1}$ must coincide with $\eta_{2}$ 
as a unique null geodesic limit curve of $\gamma_{k}$ in $U$. So, $\eta_{1}$ is a null geodesic from $p$ to $q$ 
and Case 1 leads to a contradiction.
\item[\textbf{Case 2)}]\label{case2}
$\eta_{2}\subset I^{+}(\eta_{1}(s))$, for all real values of $s\in [0,+\infty)$.\\
Consider the provided sequence $\lbrace p_{m} \rbrace$ in Part (1) of Lemma \ref{lemma order} 
by $p\preceq . . .\preceq p_{2}\preceq p_{1}$ such that $p$, $p_{n}$ are on a null geodesic passing through $p$ 
(Part (3) of Lemma \ref{lemma order} leads to a constant sequence that never happens here and for Part (2),
the proof is similar by replacing ``$+$" with ``$-$", ``$p$" with ``$q$" and ``$\eta_{1}$" with ``$\eta_{2}$"). 
From the hypothesis of this case, we immediately conclude $p\in I^{-}(q)$ 
and so there is a strictly convex normal neighborhood $U$ of $p$ such that $U\subseteq I^{-}(q)$. On the other hand, all but 
finitely many elements of $\lbrace p_{m} \rbrace$ are in $I^{-}(q)$, by the fact that $\lbrace p_{m} \rbrace$ converges
to $p$. If necessary, by selecting a suitable subsequence of $\lbrace p_{m} \rbrace$, there exists an $N_{0} \in \mathbf{N}$ such that:
\begin{equation}\label{1}
 \forall m > N_{0}\qquad  p_{m} \not\in J^{+}(p_{N_{0}}),\quad q_{m-1}, q \in  I^{+}(p_{N_{0}}).
\end{equation}
%
\begin{figure}
\vspace{0.1cm}
\hspace{3cm}
\includegraphics[width=10cm,height=11cm]{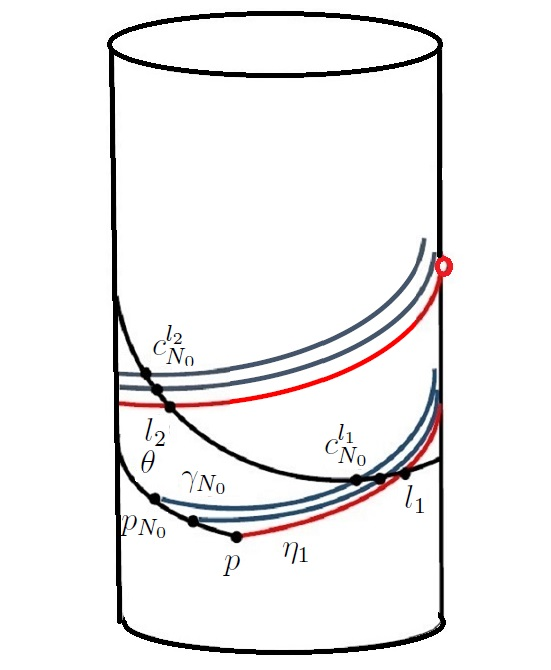}
\caption{Diagram for the proof of Theorem \ref{main}}
\label{Fig2}
\end{figure}
%
Thus, every $\gamma_{m}$ intersects $\partial J^{+}(p_{N_{0}})$ for all $m \geq N_{0}$ by \cite[Lemma 3]{Vatan}.
Let $c_{m}$ be  $\gamma_{m}(t)$ for the minimum value of $t$ in which $\gamma_{m}$
intersects $\partial J^{+}(p_{N_{0}})$. The sequence $\lbrace c_{m}\rbrace$ is a subset of 
$\partial J^{+}(p_{N_{0}})$ and by the causal simplicity condition $c_{m}\in \partial J^{+}(p_{N_{0}})\subseteq  J^{+}(p_{N_{0}})$. 
So, Corollary 4.14 in Ref. \cite{Beem(1996)} implies that there is a sequence of maximal 
null geodesic segments $\lbrace \theta_{m} \rbrace$ from $p_{N_{0}}$ to any elements of $\lbrace c_{m}\rbrace$.\\
\textbf{Claim 1:} $\lbrace c_{m}\rbrace$ has an accumulation point.
\begin{itemize}
\item[] Proof of Claim 1: Let $U(p)$ be a strictly convex normal neighborhood of $p$ such that $\overline{U}$ is compact. 
If infinitely many of $\lbrace c_{m}\rbrace$ are in $\overline{U}$, then an accumulation point $c$ is achieved. 
Otherwise, because $\partial U$ is compact, we can conclude the sequence $\lbrace \theta_{m} \rbrace$ 
has a maximal null geodesic $\theta$ with endpoint $p_{N_{0}}$ as a limit curve. 
We must prove $\lbrace c_{m}\rbrace$ has an accumulation point $c$ on $\theta$. 
In two--dimensional spacetimes, there are only two null geodesics passing through 
any point and so any $\theta_{m}$ coincides with $\theta$ and
the sequence $\lbrace c_{m}\rbrace ^{\infty}_{m=N_{0}}$ is on the null geodesic 
segment $p_{N_{0}}c_{N_{0}}\subseteq \theta$ 
as a compact set (see Fig \ref{Fig2}). Therefore, the sequence 
$\lbrace c_{m}\rbrace ^{\infty}_{m=N_{0}}$ has an accumulation point $c$ such that  
$c_{N_{0}} \succeq c_{N_{0}+1} \succeq c_{N_{0}+2}\succeq ... \succeq c_{N_{0}+n}\succeq...\succeq c$.
\end{itemize}
%
\begin{figure}
\vspace{0.1cm}
\hspace{0.5cm}
\includegraphics[width=8cm,height=10cm]{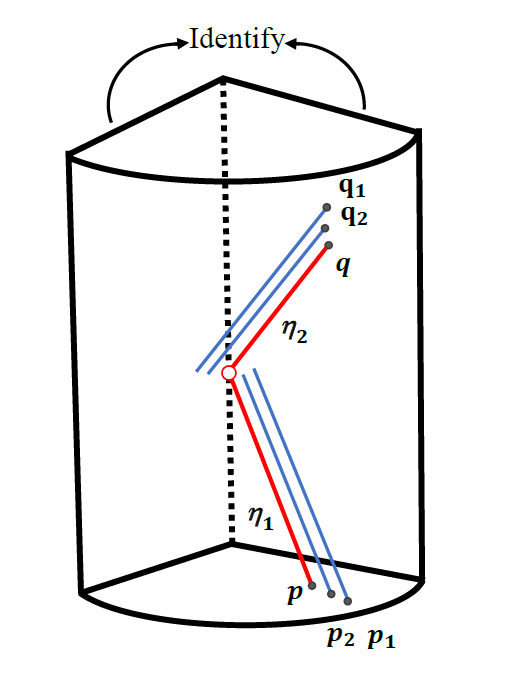}
\caption{Spacetime $M$ is causally simple but it is not null pseudoconvex. }
\label{Fig3}
\end{figure}
%
%
Let $l_{1}$ be the accumulation point of $\lbrace c_{m}\rbrace$ (i.e. $l_{1}=c$). Therefore, $l_{1}$ is on $\theta$ and $\theta$ intersects $\eta_{1}$ 
and then enters $I^{+}(p)$ for the first time at $l_{1}$, because $c_{N_{0}} \in I^{+}(p) $ and $\partial J^{+}(p) \subseteq (\theta \cap \eta_{1}$).\\
\textbf{Claim 2:} $q \in I^{+}(c_{N_{0}})$.
\begin{itemize}
\item[] Proof of claim 2: Since $q \in I^{+}(l_{1})$, there are $k_{0}\geq 0$ such that $q \in I^{+}(c_{n})$ for $n \geq N_{0}+k_{0}$. So, 
$\theta$ intersects $\gamma _{n}$ at $c_{n}^{\prime}$ for $n \geq N_{0} + k_{0}$ and the sequence 
$c_{N_{0}+k_{0}}^{\prime} \succeq c_{N_{0}+k_{0}+1}^{\prime} \succeq ...$ converges to $l_{2}$ on $\theta \cap \eta_{1}$.
Now, $c_{N_{0}} \preceq l_{2}$ and $l_{2} \prec q$ and so $c_{N_{0}} \prec q$. This means $q \in I^{+}(c_{N_{0}})$.
\end{itemize} 
Now, we start the same process. In the first step, 
label $c_{N_{0}}=c_{N_{0}}^{l_{1}}, c_{N_{0}+1}=c_{N_{0}+1}^{l_{1}},...$. Replace 
$\lbrace c_{m}^{l_{1}}\rbrace ^{\infty}_{m=N_{0}}$ and $l_{1}$ with $\lbrace p_{m} \rbrace ^{\infty}_{m=N_{0}}$
and $p$, respectively (introduced at the beginning of this case) and find $l_{2}, c_{N_{0}}^{l_{2}}, c_{N_{0}+1}^{l_{2}},...$ and repeat the process. 
At step n, we can similarly show that $q \in I^{+}(c_{N_{0}+n})$ and therefore these steps don't stop. 
So, there are infinite points $p=l_{0} \preceq l_{1} \preceq l_{2} \preceq ... \preceq l_{n} \preceq ...$ on $\eta_{1}$ and 
infinite points $c_{N_{0}}^{l_{1}} \preceq c_{N_{0}}^{l_{2}} \preceq c_{N_{0}}^{l_{3}} \preceq ... \preceq c_{N_{0}}^{l_{n}} \preceq ...$ 
on $\gamma_{N_{0}}$ such that at each step, $\eta_{1}$ and $\theta$ cut each other in $l_{n}$ and also
$\gamma_{N_{0}}$ and $\theta$ cut each other in $c_{N_{0}}^{l_{n}}$. The segments $l_{0}l_{1}, l_{1}l_{2}, l_{2}l_{3},..., l_{n-1}l_{n},... $ and 
$c_{N_{0}}^{l_{1}}c_{N_{0}}^{l_{2}}, c_{N_{0}}^{l_{2}}c_{N_{0}}^{l_{3}}, ... ,c_{N_{0}}^{l_{n-1}}c_{N_{0}}^{l_{n}} , ...$ 
are maximal null geodesics. But, this is impossible because $\gamma_{N_{0}}$ is compact and don't admit infinitely many cut-points. Therefore,
Case 2 leads to a contradiction.\\
%
%
%
%
\end{itemize}
\end{proof}
\begin{remark}
We note that Theorem \ref{main} is not true for three-dimensional spacetimes because the null pseudoconvexity and strong causality lift to 
Lorentzian covers but, there are three-dimensional spacetimes which show that causal simplicity doesn't lift \cite{Costa, Hedicke, Schinner}. 
So, we immediately conclude that these causally simple spacetimes, \cite[Example 2.3.]{Costa} and \cite[Theorem 2.7.]{Hedicke}, are not null pseudoconvex as we illustrate Example 2.3. of Ref. \cite{Costa} in Figure \ref{Fig3}. 
Therefore, the causal simplicity doesn't imply the null pseudoconvexity in three-dimensional spacetimes. 
\end{remark}
%
%
The study of the topology and the geometry of the space of null geodesics $\mathcal{N}$ gives a new approach to
considerations of causal structures of a spacetime $M$. Assuming that $M$ is strongly causal, we ensure that $\mathcal{N}$ possesses 
a differentiable structure \cite[Theorem 1]{Low(2001)}. In order for $\mathcal{N}$ to be a manifold, it is 
necessary that it be Hausdorff. For more general spacetimes, $\mathcal{N}$ may fail to be Hausdorff, as 
the interesting example of the plane wave spacetime considered by Penrose \cite{Low(2001)} shows. Low shows by 
the following theorem in Ref. \cite{Low(1990)} that the null pseudoconvexity condition of $M$ corresponds to the Hausdorffness of $\mathcal{N}$.
He describes an open problem in Ref. \cite{Low(2001)} to find necessary and sufficient causality conditions of $M$ to ensure 
that $\mathcal{N}$ is Hausdorff. Recently, Bautista, Ibort, Lafuente, and Low studied in detail the space of light rays 
and also discussed to place this necessary and sufficient property within the causal ladder, and finally 
conjectured that strongly causal null pseudoconvex spacetimes are causally simple \cite{Bautista(2017)} that 
Theorem \ref{causally simple-null pseudo} implies this conjecture.
Here, Theorems \ref{causally simple-null pseudo}, \ref{main}, and \ref{Low} 
can be applied to solve this problem in two-dimensional spacetimes, see Corollary \ref{cor1}.
%
\begin{theorem}\label{Low}
\cite{Low(1990)} Let $M$ be a strongly causal spacetime. Then the following conditions are equivalent:
\begin{itemize}
\item[1)] $M$ is null pseudoconvex.
\item[2)] The space of null geodesics, $\mathcal{N}$, is Hausdorff.
\end{itemize}
\end{theorem}
\begin{corollary}\label{cor1}
Let $\mathcal{N}$ be the space of null geodesics of a two-dimensional spacetime strongly causal $M$. Then $M$ is causally simple 
if and only if $\mathcal{N}$ is Hausdorff.
\end{corollary}
%
%
%
\subsection{Naked singularities}                                                                                     %
%
%
%
Recently, two types of naked singularities have been introduced by Borjian and Bahrampour in Ref. \cite{Borjian} 
and the relationships between the presence of each of these naked singularities in $M$, and failure of the Hausdorff 
property for $\mathcal{N}$ have been investigated. 
A spacetime $M$ is said to be a \textit{nakedly singular future boundary} if it contains some point $p$ and some future endless null geodesic 
$\Gamma$ such that $\Gamma \subseteq \partial I^{-}(p)$ and for each $q\in I^{-}(p)$, $\Gamma \cap \partial I^{-}(q)=\emptyset $. 
Also, a \textit{nakedly singular past boundary} is defined similarly by replacing $``+"$ with $``-"$.\\
In fact, Borjian and Bahrampour state a conjecture that says ``if a strongly causal spacetime 
$M$ is a nakedly singular future boundary or a nakedly singular past boundary, then $\mathcal{N}$ is non-Hausdorff". 
In this section, we prove the conjecture and show that the presence of one of these types of naked singularities implies 
the failure of the Hausdorff condition. Aso, the results of the proviose section solve the converse of this conjecture.
%
%
\begin{figure}
\vspace{0.1cm}
\hspace{3cm}
\includegraphics[width=10cm,height=8cm]{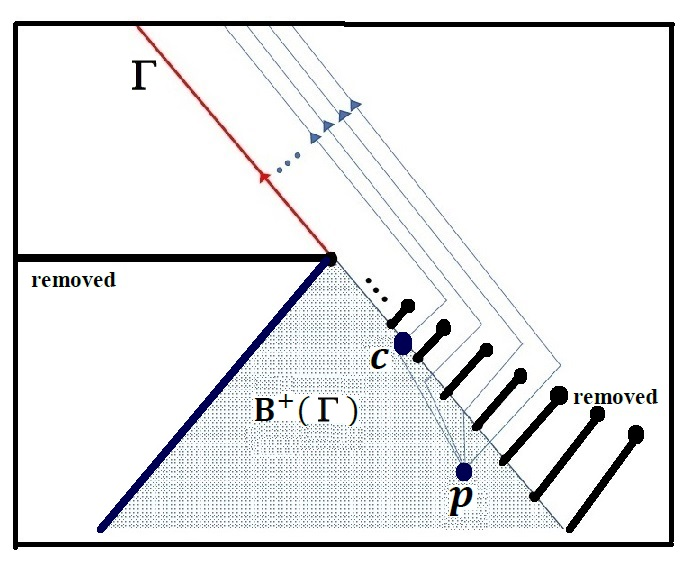}
\caption{In this non-causally continuous spacetime $B^{+}(\Gamma)$ is not closed. In fact, $c\in \partial B^{+}(\Gamma)$ but 
$c \not\in B^{+}(\Gamma)$.}
\label{Fig4}
\end{figure}
%
%
\begin{proposition}\label{property}
Let $\Gamma$ be a past (future) endless null geodesic and $B^{+}(\Gamma)=\lbrace q\in M \vert \Gamma \subseteq \partial I^{+}(q) \rbrace$ 
$(\: B^{-}(\Gamma)=\lbrace q\in M \vert \Gamma \subseteq \partial I^{-}(q) \rbrace \:).$ 
Then $B^{+}(\Gamma)$ $(\: B^{-}(\Gamma)  \:)$ is a causally convex set. Moreover, 
$B^{+}(\Gamma)$ $(\: B^{-}(\Gamma)  \:)$ is closed, if $M$ is causally continuous.
\end{proposition}
\begin{proof}
Let $\Gamma$ be a past endless null geodesic, $q \in M$ and $r, s \in B^{+}(\Gamma)$ such that $r \preceq q \preceq s$. 
We show that $q \in B^{+}(\Gamma)$. $q \preceq s$ implies that 
$\Gamma \subseteq \partial I^{+}(s) \subseteq \overline{ I^{+}(q)}$. It is sufficient to show that $\Gamma \cap \partial I^{+}(q)=\emptyset$.
On the contrary, assume that $q_{0}\in (\Gamma \cap \partial I^{+}(q))$. So, $r \preceq q \prec q_{0}$ and it implies $r \prec q_{0}$ 
(see \cite[Theorem 2.24]{Min(2019)}). Therefore, $q_{0}\in (I^{+}(r) \cap \partial I^{+}(r))$ and this is a contradiction. Thus, $q \in B^{+}(\Gamma)$.\\
Now, Let $M$ be causally continuous, and let $c_{n}\in B^{+}(\Gamma)$ and $ c_{n} \longrightarrow c$. We show that $c \in B^{+}(\Gamma)$. 
On the contrary, assume that $\Gamma \nsubseteq \partial I^{+}(c)$. There are two cases:\\
Case 1: $\exists$ $q_{0}\in (\Gamma \cap I^{+}(c))$.\\
In this case, $c \in I^{-}(q_{0})$ and since $I^{-}$ is inner continuous, there is an open neighborhood $U(c)$ of $c$ such that 
$U(c) \subseteq I^{-}(q_{0})$. So, there exists $N_{0}$ such that $c_{n}\in U(c) \subseteq I^{-}(q_{0})$ for any $n\geq N_{0}$. Therefore, 
$q_{0} \in I^{+}(c_{N_{0}})$ but $c_{n}\in B^{+}(\Gamma)$ and this is a contradiction.\\
Case 2: $\exists$ $q_{0}\in (\Gamma \cap (M \setminus \overline{I^{+}(c)}))$.\\
By assumption, since $I^{+}$ is outer continuous, there is an open neighborhood $U(c)$ of $c$ such that 
$q_{0}\in (M \setminus \overline{I^{+}(q)})$ for each $q \in U(c)$, especially for some $c_{N_{1}}\in U(c)$ but $c_{N_{1}}\in B^{+}(\Gamma)$ and this is a contradiction. Similarly, one can prove that $B^{-}(\Gamma)$ is causally convex and closed.
\end{proof}
Figure \ref{Fig4} shows that the causal continuity condition in Proposition \ref{property} is necessary.
The following proposition is proved by Beem and Krolak (see \cite [Proposition 1]{Beem(1992)}).
\begin{proposition}\label{Beem}
Assume $(M,g)$ is distinguishing, but not causally simple. Then $\exists$ points $x_{1}$ and $x_{2}$ in $M$ such that $x_{1}$ has a future
inextendible maximal null geodesic ray in $\partial I^{-}(x_{1})$ and $x_{2}$ has a past inextendible maximal null geodesic ray in $\partial I^{+}(x_{2})$.\\
Conversely, assume $(M,g)$ has an $x_{1}$ (resp. $x_{2}$) such that $\partial I^{-}(x_{1})$ has a future inextendible maximal null geodesic ray
[resp. $\partial I^{+}(x_{2})$ has a past inextendible maximal null geodesic ray]; then $(M,g)$ is not causally simple.
\end{proposition}
%
\begin{figure}
\vspace{0.1cm}
\hspace{3cm}
\includegraphics[width=10cm,height=9cm]{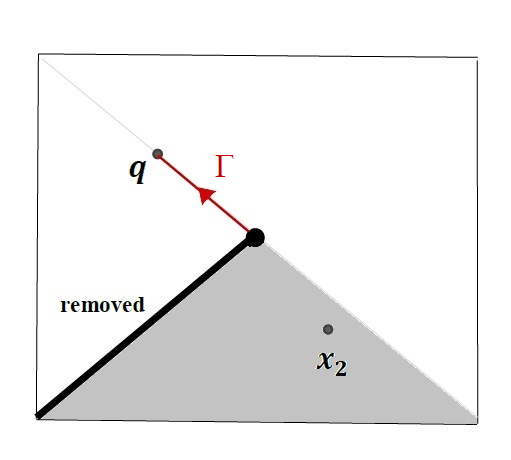}
\caption{This spacetime is the two-dimensional Minkowski space with a half-line of a null geodesic removed. 
All points in the hatched region can be chosen to be $x_{2}$ but no point in the spacetime can be $x_{1}$ in Proposition \ref{Beem}. 
The maximal past inextendible null geodesic ray $\Gamma$ is a subset of $\partial J^{+}(x_{2})$ and $J^{+}(x_{2})$ is not closed for all 
points $x_{2}$ in the hatched region but there is no maximal future inextendible null geodesic ray in $\partial J^{-}(p)$ and 
$J^{-}(p)$ is closed for all points $p\in M$. In the proof of Proposition \ref{Beem}, the invalid argument 
``we may take $q$ to be the $x_{1}$" is used.}
\label{Fig5}
\end{figure}
%
Figure \ref{Fig5} shows that this proposition is not valid and it is required to have a minor modification. 
Now, we provide a corrected version of this proposition as follows:
\begin{proposition}\label{Vatan-Beem}
Assume $(M,g)$ is distinguishing. Then $M$ is not causally simple if and only if there exists a point $x_{1} \in M$ such that $x_{1}$ has a future
inextendible maximal null geodesic ray in $\partial I^{-}(x_{1})$ or a point $x_{2} \in M$ such that $x_{2}$ has a 
past inextendible maximal null geodesic ray in $\partial I^{+}(x_{2})$.
\end{proposition}
\begin{proof}
It is sufficient to remove the following sentences in the proof of \cite[Part (I) of Proposition 1]{Beem(1992)}:\\
``The same type of argument shows that $\exists$ a maximal future inextendible null geodesic ray that starts at $x_{2}$ and 
fails to reach $q$. Thus we may take $q$ to be the $x_{1}$."\\
Instead, replace it with the following sentence:\\
``Also, if there is a point $x_{1}$ such that $J^{-}(x_{1})$ is not closed, then a similar proof shows the existence of a maximal future inextendible 
null geodesic ray in $\partial I^{-}(x_{1})$."
\end{proof}
A spacetime $M$ is said to be past (future) reflecting at $q$ in $M$ if for all $p$ in $M$
$$I^{+}(q)\subseteq I^{+}(p) \Rightarrow I^{-}(p)\subseteq I^{-}(q),$$
$$ ( I^{-}(q)\subseteq I^{-}(p) \Rightarrow I^{+}(p)\subseteq I^{+}(q) ) $$
and is said to be reflecting at $q$ if it satisfies both conditions. The spacetime is said to be reflecting if it is reflecting at all points.
It is shown that the reflectivity and the causal continuity conditions are equivalent (see \cite[Definition 4.9]{Min(2019)}). 
\begin{proposition}\label{f-pnb}
Let $(M,g)$ be a reflecting spacetime and let there exists a past (future) inextendible maximal null geodesic ray $\Gamma$ 
such that $\Gamma \subseteq \partial I^{+}(q)$ ($\Gamma \subseteq \partial I^{-}(q)$), for some $q \in M$. 
Then the following statements are true:
\begin{itemize}
\item[(I)] $M$ is a nakedly singular past (future) boundary at $p$, for all points $p \in B^{+}(\Gamma)$ ($p \in B^{-}(\Gamma)$).
\item[(II)] $int(B^{+}(\Gamma))=\emptyset$ and $\partial B^{+}(\Gamma)=B^{+}(\Gamma)$ ($int(B^{-}(\Gamma))=\emptyset$ and $\partial B^{-}(\Gamma)=B^{-}(\Gamma)$).
\end{itemize}   
\end{proposition}
\begin{proof}
By hypothesis, $q \in B^{+}(\Gamma)\not = \emptyset$. On the contrary, assume that $M$ is not a nakedly singular 
past (future) boundary at some point $p \in B^{+}(\Gamma)$.
This means that $\exists w \in I^{+}(p)$ and $v \in \Gamma \cap \partial I^{+}(w)$. Now, there are two cases:
\begin{itemize}
\item[] Case 1: $w \in \overline{I^{-}(v)}$. Since $I^{+}(p)$ is an open set containing $w$, it intersects $I^{-}(v)$ and so
$ I^{-}(v) \cap I^{+}(p)\not = \emptyset$ that implies $v \in I^{+}(p)$. This is a contradiction because of $v \in \Gamma $ 
and $\Gamma \subseteq \partial I^{+}(p)$.
\item[] Case 2: $w \not\in \overline{I^{-}(v)}$. By reflectivity of $M$, this case 
implies $v \not\in \overline{I^{+}(w)}$ but we have $v \in \partial I^{+}(w) \subseteq\overline{I^{+}(w)}$,
which is a contradiction.
\end{itemize}
Therefore, $M$ is a nakedly singular past (future) boundary at all points of $ B^{+}(\Gamma)$.\\
For the proof of the second statement, let $p\in int(B^{+}(\Gamma))\not = \emptyset$. So, $I^{+}(p) \cap B^{+}(\Gamma)$ is a non-empty set 
and we assume that $w \in I^{+}(p) \cap B^{+}(\Gamma)$. Therefore, we have $w \in I^{+}(p)$ and 
$\exists v \in \Gamma \cap \partial I^{+}(w)$. Now, there are precisely the same as the above two cases, which leads to a contradiction.
\end{proof}
Propositions \ref{Vatan-Beem} and \ref{f-pnb} immediately imply the following results:
\begin{corollary}\label{cor2}
Let $(M,g)$ be a causally continuous spacetime. Then $M$ is not causally simple if and only if $M$ is a nakedly singular future boundary or 
nakedly singular past boundary spacetime.
\end{corollary}
Now, we are ready to conclude the conjecture introduced in Ref. \cite {Borjian} by using Theorems \ref{causally simple-null pseudo},  \ref{Low} and Corollary \ref{cor2}.
\begin{corollary}\label{cor3}
Let $\mathcal{N}$ be the space of null geodesics of a causally continuous spacetime $M$. 
if $M$ is a nakedly singular future boundary or nakedly singular past boundary spacetime, then $\mathcal{N}$ is non-Hausdorff. 
\end{corollary}
Also, Theorem \ref{main} implies that the converse of Corollary \ref{cor3} is tru for two-dimensional spacetimes and Figure \ref{Fig3} refutes
the converse of Corollary \ref{cor3} for three-dimensional spacetimes.
%
%
%
%
\section{Conclusion}
The geometry of the space of null geodesics $\mathcal{N}$ of a strongly causal spacetime $M$ has provided insights into many aspects 
of spacetime geometry. Although $\mathcal{N}$ is guaranteed to have a differentiable structure, it need not be Hausdorff. 
In this paper, we find a necessary and sufficient condition through spacetime causality conditions as a solution to Problem 3 
suggested in Ref. \cite{Low(2001)}: A two-dimensional spacetime $M$ is causally simple if and only if $\mathcal{N}$ is Hausdorff. This condition is weaker than 
the causal pseudoconvexity and the global hyperbolicity conditions which are equivalent to the Hausdorffness of the space 
of causal geodesics $C$ and the space of smooth endless causal curves $\mathcal{C}$ of $M$, respectively \cite{Low(1990)}.\\
The failure of the Hausdorff condition corresponds to the presence of a particular type of naked singularity. It is shown that 
$M$ is nakedly singular if and only if $\mathcal{C}$ is non-Hausdorff. Finally, by an example, we show that a proposition 
by Beem and Krolak (see \cite[Proposition 1]{Beem(1992)}) is required to have a minor modification. Then, as a result, we prove 
$\mathcal{N}$ is non-Hausdorff if $M$ is a nakedly singular future boundary or nakedly singular past boundary spacetime 
(as a solution of \cite[Conjecture 3.1]{Borjian}) and the converse of the conjecture is true only in two-dimensional spacetimes.\\
Recently in Ref. \cite[Theorem 2.7.]{Hedicke}, Hedicke and Suhr refuted Chernov's conjecture which says that every causally simple 
spacetime can be conformally embedded as an open subset into some globally hyperbolic spacetime \cite{Chernov}. 
On the other hand, any two-dimensional simply connected and causally simple space-time can be causally isomorphically 
embedded into two-Minkowski spacetime (see \cite[Theorem 3]{Vatan-Bahram}). So, the results motivate that we conjecture the following statement:
\begin{conjecture}
Every null pseudoconvex strongly causal spacetime can be conformally embedded as an open subset into some globally hyperbolic spacetime.
\end{conjecture}
%
%
%
%
%
%

%
\end{document}